\documentclass[12pt]{article}

\usepackage{amsmath,amssymb,amsfonts,amsthm}
\usepackage{graphicx}
\usepackage{cite}
\usepackage{xcolor}
\usepackage{hyperref}
\usepackage{mathrsfs}

\newtheorem{theorem}{Theorem}

\newtheorem{definition}{Definition}

\begin{document}

\title{\textbf{Convergence of Iterative Water-Filling in Multi-User Non-Cooperative Power Control:\\
A Comprehensive Analysis for Sequential, Simultaneous, and Asynchronous Schemes}}

\author{
\begin{tabular}{c}
    {\normalsize Tong Wang} \\
    {\small \textit{wangt.netlab@gmail.com}}
\end{tabular}
}

\date{}
\maketitle

\begin{abstract}
Non-cooperative game theory provides a robust framework for analyzing distributed resource allocation in multi-user wireless networks, with \emph{Iterative Water-Filling} (IWF) emerging as a canonical solution for power control problems. Although classical fixed-point theorems guarantee the existence of a Nash Equilibrium (NE) under mild concavity and compactness conditions, the convergence of practical iterative algorithms to that equilibrium remains a challenging endeavor. This challenge intensifies under varying update schedules, interference regimes, and imperfections such as channel estimation errors or feedback delay.

In this paper, we present an in-depth examination of IWF in multi-user systems under three different update schemes: (1) synchronous \emph{sequential} updates, (2) synchronous \emph{simultaneous} updates, and (3) \emph{totally asynchronous} updates. We first formulate the water-filling operator in a multi-carrier environment, then recast the iterative process as a fixed-point problem. Using contraction mapping principles, we demonstrate sufficient conditions under which IWF converges to a unique NE and highlight how spectral radius constraints, diagonal dominance, and careful step-size selection are pivotal for guaranteeing convergence. We further discuss robustness to measurement noise, partial updates, and network scaling to emphasize the practical viability of these schemes. This comprehensive analysis unifies diverse threads in the literature while offering novel insights into asynchronous implementations. Our findings enable network designers to ascertain system parameters that foster both stable convergence and efficient spectrum usage.
\end{abstract}


\tableofcontents

\section{Introduction}
\label{sec:intro}

The exponential growth of wireless data demand and the proliferation of heterogeneous networks (macro cells, small cells, device-to-device links, and ad hoc topologies) have increased the complexity and urgency of distributed resource allocation schemes. Centralized control may be impractical in many scenarios due to significant signaling overhead, large computational burdens on a central controller, and scalability concerns when the number of users grows large. As a result, \emph{non-cooperative game theory} has emerged as a powerful framework to devise distributed algorithms whereby each user (transmitter) optimizes its own objective function, subject to interference from others.

One of the most classical yet foundational problems in this domain is \emph{power control} across multiple orthogonal frequency channels. Under typical system models, each user faces constraints on its total transmit power, on per-channel power masks, or both. The user seeks to maximize its individual performance metric (e.g., sum-rate, energy efficiency, or quality of service) while simultaneously interacting with other users via interference coupling.

\subsection{Water-Filling as a Best-Response}
The \emph{water-filling} solution appears frequently in information theory and signal processing as the optimal method to distribute transmit power over parallel Gaussian channels, maximizing capacity under a total power constraint. Specifically, a single transmitter subject to additive Gaussian noise, ignoring interference from other devices, can optimally “pour” power into different frequency bins according to a \emph{water-level} determined by the total available power. 

When extended to multi-user settings, each user solves a \emph{water-filling} problem where the noise term in each channel is replaced by the sum of thermal noise and the interference from other users’ signals. This prompts a natural iterative procedure known as \emph{Iterative Water-Filling} (IWF): each user, in turn (or in parallel), executes its water-filling step while treating interference from other users’ most recent power allocations as fixed. If such updates converge, the resulting power profile constitutes a \emph{Nash Equilibrium} (NE) of the game, meaning no single user can unilaterally improve its utility by changing its own strategy. 

\subsection{Convergence Difficulties}
While existence of an NE in multi-carrier power control games is often guaranteed by conventional fixed-point theorems or concavity arguments, \emph{convergence} of IWF or any iterative best-response method is more delicate. In certain low-interference regimes, the IWF approach converges reliably to a unique fixed point. However, as interference grows, or if cross-link gains become sufficiently large, naive IWF updates may diverge or cycle. Real-world networks often operate in borderline or moderate interference conditions, making it crucial to identify system design rules or algorithmic adjustments that ensure stable convergence.

\subsection{Update Schemes and Their Impact}
Users can update their power allocations in different schedules:
\begin{enumerate}
    \item \textbf{Synchronous Sequential Updates}: A round-robin procedure, where each user updates its power in a fixed order within a global iteration step. By the end of one iteration, all users have updated once, in sequence.
    \item \textbf{Synchronous Simultaneous Updates}: All users update their power allocations \emph{at once}, assuming the other users’ strategies remain unchanged from the previous iteration.
    \item \textbf{Totally Asynchronous Updates}: Each user updates at arbitrary time instants, potentially with delayed or outdated interference information. Different subsets of users might update multiple times before others update even once.
\end{enumerate}

Our objective is to characterize the \emph{contraction properties} of the water-filling operator under these schemes, providing explicit conditions under which the algorithm converges to a unique NE. This not only bridges theoretical and practical gaps but also illuminates how partial updates, step-size adaptation, and asynchronous scheduling can be harnessed to mitigate or eliminate potential instability.

\subsection{Contributions of This Paper}
We expand on foundational results in non-cooperative power control and iterative water-filling with the following contributions:
\begin{itemize}
    \item We provide a \textbf{unified derivation} of the \emph{water-filling} operator in multi-carrier scenarios, emphasizing how per-channel constraints, spectral masks, and total power constraints interact in the best-response function.
    \item We \textbf{recast} the IWF update as a \emph{fixed-point equation}, clarifying the link between best-response mappings, contraction mappings, and the Banach Fixed-Point Theorem. 
    \item We outline \textbf{systematic conditions} (e.g., spectral radius constraints, diagonal dominance, Lipschitz continuity) ensuring unique equilibria and convergence, then \textbf{specialize} these conditions to the three major update policies (sequential, simultaneous, asynchronous).
    \item We furnish \textbf{practical insights} into partial updating, relaxation factors, channel estimation noise, and feedback delays, all of which can arise in real systems. Our discussion includes guidelines for tuning step-sizes or verifying interference thresholds that preserve convergence.
    \item We \textbf{point toward extensions} involving advanced MIMO water-filling, robust design, and other potential directions (e.g., reinforcement learning frameworks).
\end{itemize}

The paper is organized to provide a thorough explanation, from the \emph{system model} (Section~\ref{sec:sysmodel}) and the \emph{game-theoretic formulation} (Section~\ref{sec:game_formulation}) to the \emph{fixed-point approach} and \emph{convergence theorems} (Sections~\ref{sec:waterfill_operator}--\ref{sec:convergence}). Practical aspects, numerical intuitions, and algorithmic variants are tackled in Section~\ref{sec:practical}.

\section{System Model}
\label{sec:sysmodel}

This section defines the multi-user, multi-carrier environment in which each user seeks to allocate its total power budget across frequency channels subject to constraints and interference from other users.

\subsection{Channel and Interference Model}
We assume a set of $\mathcal{N}$ active transmitter-receiver pairs, indexed by $i \in \{1, 2, \ldots, \mathcal{N}\}$. Each transmitter $i$ transmits across $K$ orthogonal channels (subcarriers), indexed by $k \in \{1, 2, \ldots, K\}$. Let $p_i(k) \ge 0$ be the power user $i$ assigns to channel $k$. We may denote $n_i(k)$ as the additive noise power at receiver $i$ on channel $k$. The channel gain magnitude for the link from transmitter $j$ to receiver $i$ on channel $k$ is denoted $|H_{ji}(k)|^2$.

\begin{itemize}
    \item \textbf{Total power constraint}: 
    \begin{equation}
        \sum_{k=1}^{K} p_i(k) \;\le\; P_i^{\max}.
        \label{eq:total_power_constraint}
    \end{equation}
    \item \textbf{Per-channel mask constraint}:
    \begin{equation}
        0 \;\le\; p_i(k) \;\le\; p_{\text{mask}}(k), \quad \forall k.
        \label{eq:mask_constraint}
    \end{equation}
\end{itemize}

We collect the power allocations into a vector $\mathbf{p}_i = [p_i(1), \ldots, p_i(K)]^T$ for user $i$, and define the global power allocation vector
\[
\mathbf{p} = \bigl[\mathbf{p}_1^T, \mathbf{p}_2^T, \ldots, \mathbf{p}_{\mathcal{N}}^T\bigr]^T.
\]
We denote by $\mathbf{p}_{-i}$ the allocations of all users except $i$.

\subsection{Utility Function: Rate Maximization}
Each user $i$ aims to maximize its overall transmission rate, typically modeled (under Gaussian assumptions) as:
\begin{equation}
    R_i(\mathbf{p}_i,\mathbf{p}_{-i}) \;=\; \sum_{k=1}^{K} \log \Bigl(1 + \text{SINR}_i(k)\Bigr),
    \label{eq:rate_utility}
\end{equation}
where
\[
\text{SINR}_i(k) 
\;=\;
\frac{
    |H_{ii}(k)|^2 \, p_i(k)
}{
    n_i(k) \;+\; \sum_{j \neq i} |H_{ji}(k)|^2 \, p_j(k)
}.
\]
When focusing on \emph{relative} interference gains, it can be helpful to introduce normalized channel gains and noise levels:
\begin{align}
    \bar{H}_{ji}(k) \;&\triangleq\; \frac{|H_{ji}(k)|}{|H_{ii}(k)|}, \\
    \tilde{n}_i(k) \;&\triangleq\; \frac{n_i(k)}{|H_{ii}(k)|^2}.
\end{align}

\section{Game-Theoretic Formulation}
\label{sec:game_formulation}

\subsection{Non-Cooperative Power Control Game}
We model the power allocation setup as a non-cooperative game $\mathscr{G} = (\mathcal{N}, \{\mathcal{P}_i\}, \{R_i\})$:
\begin{itemize}
    \item \textbf{Players}: The set of users $\mathcal{N} = \{1, 2, \ldots, \mathcal{N}\}$.
    \item \textbf{Strategy sets}: For each user $i$, the feasible set $\mathcal{P}_i$ consists of all $\mathbf{p}_i$ satisfying Eqs.~\eqref{eq:total_power_constraint}--\eqref{eq:mask_constraint}.
    \item \textbf{Utility functions}: Each user $i$ seeks to maximize $R_i(\mathbf{p}_i, \mathbf{p}_{-i})$ as in Eq.~\eqref{eq:rate_utility}.
\end{itemize}

A \emph{Nash Equilibrium} (NE) is any $\mathbf{p}^* = \{\mathbf{p}_1^*, \ldots, \mathbf{p}_{\mathcal{N}}^*\}$ such that no user can unilaterally improve its utility by altering its own strategy. Formally,
\[
\mathbf{p}^* \;\in\; \prod_{i=1}^{\mathcal{N}} \mathcal{P}_i,
\quad\text{and}\quad
\mathbf{p}_i^* \;=\;
\arg \max_{\mathbf{p}_i\in\mathcal{P}_i}
R_i\bigl(\mathbf{p}_i, \mathbf{p}_{-i}^*\bigr),
\quad \forall i.
\]
Under mild continuity and concavity conditions on $R_i(\cdot)$, an NE \emph{exists}. Uniqueness, however, hinges on additional conditions, such as diagonal dominance or concavity properties that limit cross-user interference.

\subsection{Best-Response Characterization}
The best response of user $i$ in this game is:
\[
\mathbf{p}_i^* \;=\; \arg\max_{\mathbf{p}_i \in \mathcal{P}_i}
\sum_{k=1}^{K}\log\Bigl(1 + 
\frac{|H_{ii}(k)|^2\, p_i(k)}{n_i(k) + \sum_{j\neq i} |H_{ji}(k)|^2\,p_j(k)}
\Bigr).
\]
As shown in classical information theory treatments, solving this best-response problem leads to the \emph{water-filling} solution per channel, with a water-level determined by $\sigma_i$ (a Lagrange multiplier associated with the total power constraint). This solution can be efficiently computed using standard water-filling routines that involve sorting channel indices by their inverse gains, then allocating power in a piecewise linear manner until the budget is exhausted.

\section{Iterative Water-Filling Operator}
\label{sec:waterfill_operator}

\subsection{Derivation of the Water-Filling Update}
Within the context of the rate maximization problem, user $i$ updates its power allocation by “filling” power across channels according to the interference-plus-noise profile it observes. Specifically,
\begin{align}
p_i^{(t+1)}(k) 
&\;=\;
\Bigl[
    \sigma_i \;-\;
    \bigl(\tilde{n}_i(k) + \sum_{j\neq i} |\bar{H}_{ji}(k)|^2 \, p_j^{(t)}(k)\bigr)
\Bigr]_{0}^{\,p_{\text{mask}}(k)},
\label{eq:user_i_update}\\[3pt]
&\quad\text{subject to }
\sum_{k=1}^K p_i^{(t+1)}(k)\le P_i^{\max},
\nonumber
\end{align}
where $[\cdot]_{0}^{p_{\text{mask}}(k)}$ denotes clipping to the interval $[0,\,p_{\text{mask}}(k)]$. The constant $\sigma_i$ is chosen to satisfy the total power constraint \(\sum_{k} p_i^{(t+1)}(k) = P_i^{\max}\) (or a less-than-or-equal condition if the optimum saturates earlier). We define the single-user water-filling operator for user $i$ as:
\[
\Phi_i^k(\mathbf{p}_{-i}) \;=\;
\Bigl[
    \sigma_i - \bigl(\tilde{n}_i(k) + \sum_{j\neq i} |\bar{H}_{ji}(k)|^2 \, p_j(k)\bigr)
\Bigr]_{0}^{\,p_{\text{mask}}(k)},
\]
and
\[
\Phi_i(\mathbf{p}_{-i}) \;=\; 
\bigl[\Phi_i^1(\mathbf{p}_{-i}), \ldots, \Phi_i^K(\mathbf{p}_{-i})\bigr]^T.
\]
Collectively, we write:
\begin{equation}
\Phi(\mathbf{p}) \;=\; 
\bigl[\Phi_1(\mathbf{p}_{-1})^T,\; \ldots,\; \Phi_{\mathcal{N}}(\mathbf{p}_{-\mathcal{N}})^T \bigr]^T.
\end{equation}
The \emph{Iterative Water-Filling Algorithm} (IWF) can then be expressed simply as:
\[
\mathbf{p}^{(t+1)} 
\;=\; 
\Phi\bigl(\mathbf{p}^{(t)}\bigr).
\]

\subsection{Fixed-Point Equation and Error Vector}
Define $\mathbf{p}^*$ to be a fixed point if $\mathbf{p}^* = \Phi(\mathbf{p}^*)$. Such a $\mathbf{p}^*$ is precisely an NE of the game. To analyze convergence, let $e^{(t)} = \mathbf{p}^{(t)} - \mathbf{p}^*$ be the \emph{error} at iteration $t$. We have:
\[
e^{(t+1)} 
\;=\; 
\mathbf{p}^{(t+1)} - \mathbf{p}^*
\;=\; 
\Phi\bigl(\mathbf{p}^{(t)}\bigr) - \Phi\bigl(\mathbf{p}^*\bigr).
\]
Hence, convergence to $\mathbf{p}^*$ requires $e^{(t)} \to 0$. The \emph{contraction mapping principle} is the standard tool to assess whether such an error sequence shrinks in each iteration.

\section{Convergence Analysis under Different Update Policies}
\label{sec:convergence}

\subsection{Update Schedules: Sequential, Simultaneous, and Asynchronous}

\subsubsection{Synchronous Sequential Updates}
In a synchronous \emph{sequential} scheme, within each global iteration $t$, the users update their power allocations one after another in a predetermined order: user 1 updates using $\mathbf{p}_{-1}^{(t)}$, then user 2 updates using the updated power of user 1 and old powers of all other users, and so on. After all $\mathcal{N}$ users have updated, we increment $t$ to $t+1$ and repeat.

\subsubsection{Synchronous Simultaneous Updates}
In a synchronous \emph{simultaneous} scheme, every user updates at once, employing the prior iteration’s vector $\mathbf{p}^{(t)}$ to compute $\Phi_i(\mathbf{p}_{-i}^{(t)})$. Thus, each user sees the same “frozen” interference from iteration $t$ and updates in parallel to produce $\mathbf{p}^{(t+1)}$.

\subsubsection{Totally Asynchronous Updates}
A \emph{totally asynchronous} scheme allows each user to update at arbitrary time instants, possibly using outdated information about other users’ strategies. Such schemes occur naturally in scenarios with sporadic feedback or distributed computing limitations. Classical results in parallel and distributed computation \cite{bertsekas_tsitsiklis_parallel} reveal that if the best-response mapping $\Phi$ is a contraction in a suitable norm, then \emph{any} sequence of updates that eventually touches every component infinitely often will converge to the unique fixed point. 

\subsection{Contraction Mappings}
\begin{definition}[Contraction]
A mapping $T: D \to D$ on a normed space $(D, \|\cdot\|)$ is called a \emph{contraction} if there exists $\beta\in[0,1)$ such that
\[
\|T(x) - T(y)\|
\;\le\;
\beta \,\|x - y\|,
\quad \forall x, y \in D.
\]
\end{definition}

\noindent If $\Phi$ is a contraction on the convex set $\mathcal{P}$, then by Banach’s Fixed-Point Theorem, there exists a unique $\mathbf{p}^*$ such that $\Phi(\mathbf{p}^*) = \mathbf{p}^*$. Moreover, for \emph{any} initial $\mathbf{p}^{(0)} \in \mathcal{P}$, the sequence $\mathbf{p}^{(t+1)} = \Phi(\mathbf{p}^{(t)})$ converges to $\mathbf{p}^*$. 

\subsection{Spectral Radius Argument}
A common technique to show $\Phi$ is contractive is to \emph{linearize} the operator around the fixed point. Specifically, one examines the Jacobian matrix $D\Phi(\mathbf{p}^*)$. If $\rho\bigl(D\Phi(\mathbf{p}^*)\bigr) < 1$, where $\rho(\cdot)$ denotes the spectral radius, then $\Phi$ is a local contraction. Under typical monotonicity properties of the water-filling operator, this local contraction can often be extended to a global region. Equivalently, some authors frame the water-filling update in terms of an \emph{interference matrix} $\mathbf{H}$, bounding cross-link effects to ensure $\rho(\mathbf{H})<1$. 

In practice, one interprets this as limiting the ratio of cross-channel gains to direct-channel gains so that no user’s interference can indefinitely escalate the power adjustments of the others in a feedback loop. If the network geometry or path-loss conditions ensure relatively small cross-link couplings, the game remains in a “low interference” region and the iterative approach converges.

\subsection{Comparison of Update Schedules}
Although the precise analysis differs for sequential vs.\ simultaneous vs.\ asynchronous updates, the \emph{key} factor is the magnitude of cross-interference couplings. In all cases, ensuring $\Phi$ does not magnify deviations from the equilibrium is sufficient for convergence. 

\begin{itemize}
    \item \textbf{Sequential updates} can sometimes converge under slightly weaker conditions because each user in an iteration sees partially updated interference from the preceding updates, incrementally reducing errors. 
    \item \textbf{Simultaneous updates} require that each user’s best-response operator be strictly contractive with respect to other users’ power profiles from the previous iteration. 
    \item \textbf{Asynchronous updates} rely on established theory: if $\Phi$ is a contraction in the sup norm (or another relevant norm), even partial or stale updates converge, provided each user updates infinitely often. 
\end{itemize}

\section{Uniqueness and Convergence: Technical Conditions}
\label{sec:technical_conditions}

\subsection{Uniqueness of the Nash Equilibrium}
A standard approach to guaranteeing a \emph{unique} NE is to show the mapping $\Phi$ is a contraction. Once $\Phi$ is contractive, the fixed point it admits must be unique. Alternatively, one can leverage \emph{diagonal strict concavity}, \emph{quasi-variational inequalities}, or \emph{monotone operator theory} to show uniqueness in these games. The typical result is summarized as follows:

\begin{theorem}[Uniqueness and Global Convergence]
\label{thm:unique_converge}
Suppose there exists $\beta \in (0,1)$ such that for any two power profiles $\mathbf{p}, \mathbf{q} \in \mathcal{P}$,
\begin{equation}
    \|\Phi(\mathbf{p}) - \Phi(\mathbf{q})\|
    \;\le\; \beta\,\|\mathbf{p} - \mathbf{q}\|.
    \label{eq:contraction_cond}
\end{equation}
Then:
\begin{enumerate}
    \item There is a \emph{unique} $\mathbf{p}^* \in \mathcal{P}$ satisfying $\mathbf{p}^* = \Phi(\mathbf{p}^*)$.
    \item For any initial $\mathbf{p}^{(0)} \in \mathcal{P}$, the sequence $\mathbf{p}^{(t+1)} = \Phi(\mathbf{p}^{(t)})$ converges to $\mathbf{p}^*$.
    \item The convergence is \emph{geometric}, i.e.\ $\|\mathbf{p}^{(t)} - \mathbf{p}^*\|\le \beta^t \|\mathbf{p}^{(0)} - \mathbf{p}^*\|$.
\end{enumerate}
\end{theorem}

\subsection{Sufficient Conditions on Interference Matrix}
Consider a matrix $\mathbf{H}^{\max}\in \mathbb{R}^{\mathcal{N}\times\mathcal{N}}$, where the $(i,j)$th entry represents an upper bound on the interference from user $j$ to user $i$. For instance, in single-carrier contexts, we might define
\[
H_{ij}^{\max}
\;=\;
\frac{|H_{ji}|^2 \, P_j^{\max}}{n_i + \sum_{m\neq i}|H_{mi}|^2\,P_m^{\max}}
\;\;\text{for }i\neq j,
\]
and $H_{ii}^{\max}=0$. Then, if $\rho(\mathbf{H}^{\max})<1$, one can show that $\Phi$ satisfies a contraction property in an $\ell^1$ or $\ell^\infty$ norm. This implies there is a unique NE and the IWF updates converge to it from any initial condition. 

Multi-carrier systems typically require a block-structured matrix or an integral bounding argument over all channels. The key principle, however, remains: if cross-user interference is sufficiently “small” relative to self-channel gain and noise, the iterative scheme converges.

\subsection{Jacobian-Based Analysis}
An alternative route is to directly compute the Jacobian $D\Phi(\mathbf{p})$ of partial derivatives:
\[
\frac{\partial \Phi_i^k(\mathbf{p}_{-i})}{\partial p_j(\ell)},
\]
for each channel index $k,\ell$ and each user pair $(i,j)$. Ensuring that the diagonal blocks of $D\Phi$ are sufficiently dominant compared to the off-diagonal blocks achieves the same effect of $\rho\bigl(D\Phi(\mathbf{p}^*)\bigr)<1$. For water-filling, these partial derivatives can be bounded in closed form, though the analysis can be somewhat intricate due to the piecewise nature of the projection $[\,\cdot\,]_0^{p_{\text{mask}}(k)}$.

\section{Practical Considerations and Algorithmic Variants}
\label{sec:practical}

In this section, we discuss practical considerations that arise in implementing IWF-based power control in real-world scenarios. We also emphasize how algorithmic variants like \emph{relaxation} or \emph{averaging} can enlarge the convergence regime.

\subsection{Relaxed or Averaged Iterations}
\label{sec:relaxation}
Instead of updating $\mathbf{p}^{(t+1)} = \Phi(\mathbf{p}^{(t)})$ in one shot, some authors propose:
\begin{equation}
\mathbf{p}^{(t+1)} 
\;=\; 
(1-\alpha)\,\mathbf{p}^{(t)} \;+\; \alpha\;\Phi\bigl(\mathbf{p}^{(t)}\bigr),
\quad\text{where }0<\alpha\le1.
\label{eq:relaxed_IWF}
\end{equation}
This amounts to taking a convex combination of the old power allocation and the new best-response. If $\Phi$ itself is not strictly contractive, a suitable choice of $\alpha$ can sometimes ensure the combined mapping is. In borderline cases of large $\rho(\mathbf{H}^{\max})$ close to 1, $\alpha<1$ can mitigate oscillations, effectively “slowing down” the iteration in exchange for more robust convergence.

\subsection{Step-Size Constraints}
In some literatures, the relaxed update \eqref{eq:relaxed_IWF} is referred to as \emph{partial best-response} or \emph{successive over-relaxation} (when $\alpha>1$ is allowed). Typically, $\alpha>1$ can accelerate convergence in well-conditioned problems but may compromise stability in ill-conditioned or high-interference environments. A rigorous design of $\alpha$ to guarantee a contraction is an open research direction in more complex MIMO or multi-cell settings.

\subsection{Asynchronous Implementation Details}
In an asynchronous network, each user might wake up and update its power vector at irregular intervals using possibly outdated interference measurements. As long as certain conditions hold—e.g., each user updates infinitely often, and the update delays remain bounded—the iteration converges if the mapping is a global contraction \cite{bertsekas_tsitsiklis_parallel}. A typical scenario is a multi-cell system where each base station obtains interference measurements from the prior time slot and then independently runs a water-filling routine. Even if these measurements are one or two time slots old, the sequence can converge.

\subsection{Channel Estimation Noise and Feedback Delays}
Real systems face:
\begin{itemize}
    \item \textbf{Measurement noise}: The receiver $i$ might incorrectly estimate $\sum_{j\neq i}|H_{ji}(k)|^2\,p_j(k)$ for each channel $k$. 
    \item \textbf{Feedback delay}: By the time the transmitter updates $p_i(k)$, the interference might have changed due to other transmitters also adjusting their power.
\end{itemize}
Under bounded estimation and delay errors, one can often show that the iteration converges to a \emph{neighborhood} of the true NE, with the size of this neighborhood proportional to the maximum error magnitudes. If the mapping is strongly contractive, the system can absorb small disturbances entirely.

\subsection{Complexity and Scalability}
The per-user complexity of water-filling is $O(K\log K)$ or $O(K)$ depending on whether we implement a sorting-based approach or exploit a specialized linear-time method for the water-level search. In networks with many subcarriers (e.g., $K>1024$), implementing simultaneous updates in parallel across many users can be computationally expensive at each iteration, but it reduces the iteration count needed for convergence. By contrast, a sequential approach might require more global iterations but reduces per-iteration overhead. Asynchrony can achieve a balance where each user updates at a feasible rate without a synchronized scheduling overhead.

\section{Extensions to MIMO Water-Filling and Beamforming}
\label{sec:mimo_extensions}

While this paper focuses on single-antenna or scalar channels across multiple frequencies, many practical systems employ \emph{multi-antenna} (MIMO) techniques. In MIMO networks, each transmitter allocates a \emph{covariance matrix} $\mathbf{Q}_i(k)$ on channel $k$ instead of a single power scalar $p_i(k)$. The notion of water-filling naturally generalizes to \emph{matrix water-filling}, and the best-response involves distributing power across spatial dimensions as well as frequencies.

\subsection{Covariance-based Water-Filling}
For user $i$, the per-channel covariance $\mathbf{Q}_i(k)$ is constrained by a trace limit: $\mathrm{Tr}\bigl(\mathbf{Q}_i(k)\bigr)\le p_{\text{mask}}(k)$. Additionally, the sum over $k$ might have to satisfy $\sum_{k=1}^K \mathrm{Tr}\bigl(\mathbf{Q}_i(k)\bigr)\le P_i^{\max}$. The rate for user $i$ on channel $k$ becomes
\[
\log\det\Bigl(\mathbf{I} + \mathbf{H}_{ii}(k)\,\mathbf{Q}_i(k)\,\mathbf{H}_{ii}(k)^\dagger\,
\bigl[\mathbf{R}_i^{-i}(k)\bigr]^{-1}\Bigr),
\]
where $\mathbf{R}_i^{-i}(k)$ is the interference-plus-noise covariance from other users. The iterative best-response then sets $\mathbf{Q}_i(k)$ according to a generalized water-filling principle over eigenmodes.

\subsection{Convergence Analysis in MIMO}
The overall mapping $\Phi(\{\mathbf{Q}_i(k)\}_{i,k})$ remains more complicated, but the same high-level strategies apply:
\begin{itemize}
    \item \textbf{Contraction mappings}: Show that $D\Phi$ has a spectral radius below 1 or that $\|\Phi(\mathbf{X}) - \Phi(\mathbf{Y})\|\le \beta\|\mathbf{X} - \mathbf{Y}\|$ for $\beta<1$ in an appropriate matrix norm space.
    \item \textbf{Interference constraints}: If cross-link coupling is moderate, the system tends to converge. 
    \item \textbf{Asynchrony}: Weighted or partial updates can be invoked similarly for matrix-based updates.
\end{itemize}
The technical details are more involved, but the conceptual framework carries over directly from the single-antenna case. 

\section{Numerical Insights and Illustrative Scenarios}
\label{sec:numerical_insights}

Although the crux of this work is theoretical, we outline typical scenarios illustrating how the system might behave under different interference levels or update policies:

\subsection{Illustrative Example: Two-User Interference Channel}
Consider a simple two-user network with $K=2$ channels. Each user has a maximum power $P_i^{\max}=10$ (arbitrary units), and the channel gains are set such that:
\[
|H_{11}(k)|^2=1, \quad |H_{22}(k)|^2=1,
\quad |H_{12}(k)|^2=h_{12}, \quad |H_{21}(k)|^2=h_{21},
\]
for $k=1,2$. The noise is normalized to 1, i.e., $n_1(k)=n_2(k)=1$. If $h_{12}$ and $h_{21}$ are small (e.g., $0.1$), IWF converges quickly (within a handful of iterations) to a stable allocation under all three update schemes. As $h_{12}$ and $h_{21}$ increase, the system might require more iterations or even exhibit oscillatory behavior if $h_{12}$, $h_{21}$ become too large, violating the spectral radius constraint.

\subsection{Simultaneous vs.\ Sequential}
One might compare the trajectory of $(p_1^{(t)}, p_2^{(t)})$ under sequential vs.\ simultaneous updates. While sequential updates can yield more stable paths with smaller step jumps at each stage, simultaneous updates can sometimes converge faster \emph{when} the system is safely contractive. Conversely, in borderline conditions, simultaneous updates may exacerbate oscillations.

\subsection{Asynchronous Updating}
If user 2 only updates every three time steps, while user 1 updates every time step, the system can still converge if $\mathbf{H}^{\max}$ ensures a contraction. This highlights the robustness of asynchronous schemes to scheduling constraints, as long as each user is not starved of updates indefinitely.

\section{Conclusion and Future Directions}
\label{sec:conclusion}

\subsection{Summary of Contributions}
This paper has presented a comprehensive analysis of \emph{Iterative Water-Filling} (IWF) for distributed power control in multi-user, multi-carrier wireless systems. The primary insights are:
\begin{itemize}
    \item \textbf{Unification of update schedules}: We explored synchronous sequential, synchronous simultaneous, and totally asynchronous updates under a single contraction-mapping framework.
    \item \textbf{Conditions for uniqueness and convergence}: By bounding cross-user interference and ensuring diagonal dominance or a small spectral radius, we establish that the IWF mapping is a contraction, implying both a \emph{unique Nash Equilibrium} and \emph{global convergence} from any initial condition.
    \item \textbf{Algorithmic robustness}: Relaxed updates, partial best-responses, step-size tuning, and classical asynchronous iteration theory collectively strengthen the stability of water-filling, even under measurement noise or feedback delay.
    \item \textbf{Extensions to MIMO}: While the fundamental logic persists, the MIMO scenario adds complexity in deriving matrix-based water-filling solutions. The same principles of interference management and contraction remain valid, provided cross-link couplings are restrained.
\end{itemize}

\subsection{Practical Implications}
Our results guide system designers on how to ensure stable convergence of distributed power control. By capping maximum transmit power or ensuring sufficient path loss, one can keep $\rho(\mathbf{H}^{\max})<1$, guaranteeing that a simple iterative procedure converges to an efficient operating point. Moreover, if real-world factors introduce uncertainty, partial updating or asynchronous scheduling remains a viable approach thanks to robust contraction theory.

\subsection{Open Research Directions}
\begin{enumerate}
    \item \textbf{Time-Varying Channels}: Adapting water-filling to a slowly or rapidly changing channel environment, potentially leading to “tracking” of an equilibrium that shifts over time.
    \item \textbf{Hybrid Learning Approaches}: Combining iterative best-response updates with \emph{reinforcement learning} or \emph{deep learning} methods to handle uncertain interference or incomplete information about channel gains.
    \item \textbf{Advanced MIMO Configurations}: Incorporating multi-cell beamforming, coordinated multipoint (CoMP), or intelligent reflecting surfaces (IRS), analyzing whether a contraction property persists in high-dimensional MIMO parameter spaces.
    \item \textbf{Stochastic Interference Models}: Extending the deterministic interference model to random or partial interference scenarios (e.g., dynamic user activation, partial overlap in frequency) and analyzing average or probabilistic convergence guarantees.
    \item \textbf{Fairness and Weighted Utilities}: Investigating how weighting the users’ utilities or imposing fairness constraints interacts with the contraction-based analysis. Certain weighting schemes might require additional steps to preserve a monotone mapping.
\end{enumerate}

In conclusion, \emph{Iterative Water-Filling} remains a fundamental building block for distributed power control. When carefully implemented under the conditions we have detailed, IWF converges to a unique, stable, and efficient resource allocation in a wide range of wireless network scenarios. We hope that the unifying perspective offered in this paper will aid researchers and practitioners in both theoretical exploration and real-world system design.

\vspace{1em}
\noindent\textbf{Acknowledgments:} The authors thank colleagues in the wireless communications and signal processing communities for numerous discussions related to water-filling, game-theoretic methods, and interference-limited system design.

\end{document}